# Single-pass, second harmonic generation of ultrafast, higher order vector vortex beams at blue

RAVI KIRAN SARIPALLI,[1,] * ANIRBAN GHOSH,[1] N. APURV CHAITANYA,[2] AND G. K. SAMANTA[1]

[1]Photonic Sciences Lab., Physical Research Laboratory, Navrangpura, Ahmedabad 380009, Gujarat, India
[2]Tecnologico de Monterrey, Escuela de Ingeniería y Ciencias, Ave. Eugenio Garza Sada 2501, Monterrey, N.L., México, 64849
*Author e-mail address: saripark85@gmail.com

**We report a novel experimental scheme for single-pass second harmonic generation (SHG) of vector vortex beam in the blue. Using an ultrafast Ti:Sapphire laser of pulse width ~17 *fs* and a set of spiral phase plates in polarization based Mach-Zehnder interferometer (MZI) we have generated vector vortex beams of order as high as $l_p$ = 12 at an average power of 860 mW. Given the space-variant polarization of the vector vortex beam, and the dependence of nonlinear frequency conversion processes on the polarization of the interacting beams, using two contiguous bismuth borate crystals with optic axis orthogonal to each other, we have frequency-doubled the near-IR vector vortex beam into visible vector vortex beams with order as high as $l_{sh}$=24. The maximum output power of the vector vortex beam of order, $l_{sh}$ =2 is measured be as high as 20.5 mW at a single-pass SHG efficiency of 2.4 %. Controlling the temporal delay in the MZI, we have preserved the vector vortex nature of the beams at both pump and frequency-doubled beams at ultrafast timescales. The measurement on mode purity confirms the generation of high quality vector vortex beams at pump and SHG wavelengths. The generic experimental scheme can be used to generate vector vortex beams across the electromagnetic spectrum.**

Light beams carry angular momentum in the form of both spin angular momentum (SAM) and orbital angular momentum (OAM). While the SAM is associated to the polarization of the light beam, the OAM arises from the helicity in the Poynting vector of the beams. The optical beams carrying OAM, commonly represented as vortex beams, have azimuthal phase pattern represented by *exp(ilφ)*, leading to phase singularity and doughnut shaped intensity profile [1]. Here, *l*, is known as the topological charge or order of the vortex beam carrying OAM of *lℏ* per photon, and *φ* is the azimuthal angle. Like the scalar vortex beams, having homogeneous polarization in the transverse plane, the vortex beams with space-variant polarization in the beam transverse plane known as *vector vortex* beams have attracted a great interest in recent times.

The vector vortex beams can be represented by mapping the total angular momentum onto a higher-order Poincaré (HOP) sphere [2, 3, 4]. Due to the presence of both SAM and OAM, the vector vortex beams have found applications in material processing [5], fiber mode selection [6], optical trapping [7], photon entanglement [8], microscopy [9], and space and mode division multiplexing [10]. The vector vortex beams are generated through the manipulation of polarization and spatial mode of a laser beam using spatially varying retarders [11], q-plates and phase retarders [12], nematic spatial light modulator [13], or by interferometric coaxial superposition of orthogonally polarized vortex beams of same topological charges but with opposite sign [14]. Unfortunately, the material dispersion, optical damage threshold and the wavelength coverage of optical elements modulating the polarization and spatial mode restrict the generation of vector vortex beams to a particular wavelength range and output power.

Nonlinear frequency conversion processes have evolved as a simple way of generating optical radiation across the electromagnetic spectrum without degrading the spatial structure of the input laser. As a result, significant progress has been made, in recent times, to generate scalar vortex beams using frequency conversion processes [15-17]. Therefore, it is imperative to use nonlinear processes to generate vector vortex beams across the electromagnetic spectrum. However, the presence of varying polarization in the transverse plane of the vector vortex beam and the polarization dependence of phase-matching of nonlinear frequency conversion processes restrict the nonlinear generation of vector vortex beams. A type-II second harmonic generation (SHG) scheme of such beams would be possible as the scheme allows interaction of orthogonal polarizations. However, at ultrafast timescale these schemes do not generate a vector vortex in the second harmonic (SH) [18-20]. Recently, we have demonstrated the direct transfer of vector vortex beam nature of the pump beam to the down converted photons in a polarization Sagnac interferometer [8]. Using similar technique, efforts have been made to generate vector beams in the SH by splitting the orthogonal polarizations modes of the pump vector vortex beam and recombining the frequency-doubled modes through the manipulation of polarization [21]. Here we report, a simple experimental scheme to generate vector vortex beams at new wavelength. Using single-pass SHG of ultrafast vector vortex beam at 810 nm in two crystals sandwiched with their optic axes oriented in orthogonal planes, we have generated vector vortex beams of orders as high as *l*=24 at 405 nm and output power in excess of 20 mW.

The schematic of the experimental setup is shown in Fig. 1. A Ti:Sapphire laser delivering output pulses of average power of 1 W at 810 nm with a spectral width of ~55 nm and pulse-width of ~17 fs at a repetition rate of 80 MHz is used as the fundamental source. Three spiral phase plates (SPPs) having phase variation corresponding to the vortex orders, *l*= 1, 2 and 3 in combination with a vortex doubler setup consisting of polarizing beam splitter cube (PBS1), quarter-wave plate (QWP1) and mirror, M1 are used to convert Gaussian beam into optical vortices of orders *l* = 1-12. The working principle of the vortex-doubler has been explained in our previous work [16]. The half-wave plate (HWP1) before the vortex doubler controls the laser power. Using a polarization based MZI consisting of HWP2, PBS2, PBS3 and mirrors M2-M8, we have converted scalar vortex beams into classical non-separable states (vector vortex beam) in OAM and polarization degrees-of-freedom (DoFs). A delay line comprised with mirrors, M4 and M5, is introduced in one of the arms to ensure temporal overlap between the two beams from the MZI. To reduce the temporal broadening of the laser due to the presence of large number of optical elements in beam path, a pair of chirp mirror (GDD ~ -175$fs^2$ per reflection) is used before the nonlinear crystal. Using the lens, L1, the vector vortex beam is focused to the nonlinear crystal and subsequently collimated using lens, L2. A dual-crystal scheme having two contiguous BIBO (bismuth borate) crystals each of 0.6-mm thickness and 10 × 10 mm$^2$ in aperture with their optic axes aligned in perpendicular planes is used as the nonlinear crystal. Both the crystals are cut with, θ = 151.7° (φ = 90°) in optical *yz*-plane for perfect phase-matching of type-I (*e* + *e*→ *o*), frequency doubling of 810 nm into 405 nm. The separator, S, extracts SH beam from the fundamental. The fundamental and SH beams are characterized using the polarization projections technique [7] comprised with QWP3, HWP4, PBS5 and QWP2, HWP3, PBS4 respectively. The states representing different points on the higher-order Poincaré sphere are generated by rotation of a QWP and HWP about their optic axis.

The electric field of the output state of MZI is a superposition of a vertically polarized beam ($E^V$) of vortex order $l_p$ and a horizontally polarized beam ($E^H$) of the same vortex order, but opposite sign -$l_p$, and can be represented as, $E_p = 1/\sqrt{2}(|E_p^H, -l_p\rangle + |E_p^V, l_p\rangle)$. To verify the generation of non-separable states of the pump and SH beams, the intensity profiles of both the beams projected in different polarization states were studied. The intensity distribution of the beams measured without any polarization projection is shown in Fig. 2(a). To verify the non-separability, measurement in one DoF influencing the outcome of the measurement in other DoF, we recorded the intensity distribution of the pump beam of order |$l_p$| = 3, for different polarization states, horizontal (H), vertical (V), anti-diagonal, (A), diagonal (D), right circular (R), and left circular (L). The results are shown in Fig. 2(a). As evident from Fig. 2(a), the projection of the pump beam in H and V polarizations result into vortices of order, $l_p$, and −$l_p$, respectively. However, the projection at A, D, L and R polarizations result ring lattice structure containing 2$l_p$ =6 number of petals at different orientations due to the superposition of two vortices of same order but in opposite sense of rotation. All these projected intensity distributions represent different points on the higher order Poincaré sphere [2, 3].

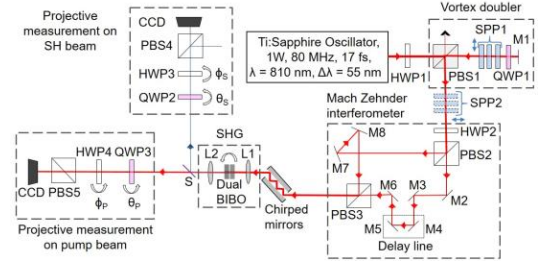

Fig. 1. Schematic experimental setup. HWP1-4: half wave plates; PBS1-5: polarizing beam splitter cubes; SPP1 and SPP2: spiral phase plates; QWP1-3: quarter wave plates; M1-8: mirrors; L1-2: lenses; S: harmonic separator; CCD: charge coupled device.

Focusing the pump vortex beam of power 860 mW using the lens, L1, of focal length *f*= 25 mm at the centre of the dual-crystal, we have verified the SH vector vortex beam by recording the intensity distribution of the beam for different polarization states with the results shown in Fig. 2(b). As evident from Fig. 2(b), in absence of any polarization projection the SH beam has a doughnut intensity profile similar to the pump beam. While the beam has doughnut intensity profile for H and V polarization states, we observe ring lattice structure of different orientation for the beam projection at D, A, L, and R polarizations represented by different points in the higher order Poincaré sphere. Such observation clearly confirms the non-separability or the vector vortex nature of the SH beam generated through nonlinear process. The number of lobes of the SH beam at diagonal projection determines the vortex order to be |$l_{sh}$|=2 x |$l_p$|=6, confirming the OAM conservation in SHG process.

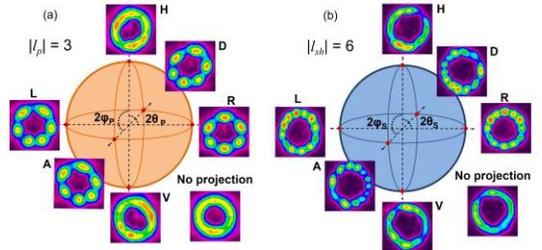

Fig. 2. Intensity profiles of the (a) pump beam, |$l_p$|=3, and corresponding (b) SHG beam, |$l_{sh}$|=6 depending on the projection of the beams at different polarization states, H, D, R, V, A and L.

The nonlinear generation of vector vortex beams in SHG process can be understood as follows. Due to the orthogonal orientation of the optic axes of the contiguous dual-BIBO crystals, the horizontal (vertical) component of the pump vector vortex beam is frequency-doubled into SH beam of vertical (horizontal) polarization. On the other hand, the OAM conservation of SHG process doubles the OAM of the SH beam keeping the vortex sign same as the pump. As a result, the generated SH beam constitutes vortex beams of orthogonal polarizations (H and V) with same order (|$l_{sh}$|) but in opposite sense of rotation (sign, ±) in the form of a vector vortex beam, $E_{sh} = 1/\sqrt{2}(|E_{sh}^V, -l_{sh}\rangle + e^{i\varphi}|E_{sh}^H, l_{sh}\rangle)$. Due to the material dispersion, the orthogonal polarization states of both pump and SH beams travel with different phase velocities resulting in a time delay (phase, $\varphi$) between the components of the vector vortex beam. Such temporal mismatch between the orthogonal components, especially, at ultrafast time scale, destroys the non-separability of the OAM and polarization DoFs and transforms the vector vortex beam into superposed scalar vortex beams. The use of temporal delay line between the orthogonal components of the pump vector vortex beams inside the MZI maintains the non-separability of the SH vector vortex beams.

We further studied the nonlinear generation of vector vortex beams of different orders. Using three SPPs corresponding to the vortex orders, $l$= 1, 2 and 3 in different combinations along with the vortex doubler, we have generated pump vector vortex beams of different orders and subsequently frequency-doubled. The intensity distributions without any polarization projection are shown in the first and third column for the pump and SH vector vortex beams respectively. and for the diagonal projection of the pump and SH beams are shown in Fig. 3. As evident from second column of Fig. 3, the $2l_p$ number of petals of the ring lattice structure confirms the pump vector vortex beam of orders, $|l_p|$ = 6, 7 and 12. Similarly, counting the number of petals of the ring lattice of the SH vector vortex beams, as shown in the fourth column of Fig. 3, we confirm the order the SH vector vortex beams to be $|l_{sh}|$ = 2 x $|l_p|$ = 12, 14 and 24, twice the order of the pump vector vortex beams. At higher vortex orders, we observe asymmetry and disintegration due to the perturbation, anisotropy, and also the walk-off effect arising from the birefringence of the nonlinear media [22]. Using periodic poled crystals (quasi phase matching) would help in improving the quality of the generated vector vortex beams at higher orders.

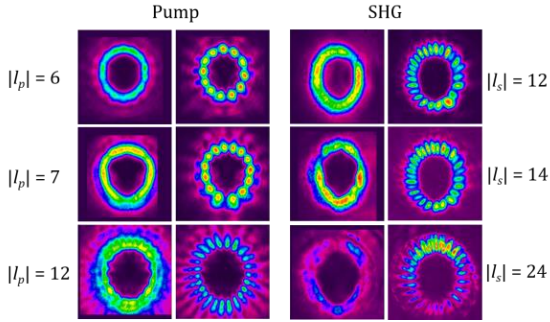

Fig. 3. The first and third colums show intensity distribution of pump vector vortex beam of orders, $|l_p|$= 6, 7, 12 without and with diagonal polarization projection respectively. Third and fourth columns show intensity distribution of corresponding SHG beams of order $|l_{sh}|$= $2|l_p|$ without and with diagonal polarization projection.

We have also verified the control of weightage of the modes of the vector vortex beam. The output state of the SH beam can be written as $E_{sh} = \alpha_1 |E_{sh}^V, -l_{sh}\rangle + \alpha_2 |E_{sh}^H, l_{sh}\rangle$, where, $\alpha_1$ and $\alpha_2$ are the relative weightages of the states at poles of the higher order Poincaré sphere. Pumping the crystal with vector vortex beam of order, $|l_p|$ = 1, we have rotated the HWP2 by an angle $\beta_p$ and measured the intensity of the vertical and horizontal components of SH beam with the results shown in Fig. 4(a). As evident from Fig. 4(a), the relative intensity of the modes varying from $\alpha_1^2$ = 1, $\alpha_2^2$ = 0 to $\alpha_1^2$ = 0, $\alpha_2^2$ = 1 with the rotation of HWP2 from $\beta_p$ = -22.5° to 22.5° with $\alpha_1^2$= $\alpha_2^2$ =0.5 at $\beta_p$=0° in close agreement with the theoretical results (solid lines). Controlling the weightage of the modes we have further measured the mode purity of the vector vortex beam at both pump and SH wavelengths using mode projection technique [23, 24]. As evident from Fig. 4(b), the pump vector vortex of order $|l_p|$ = 1 has two distinct peaks at OAM mode of $l$ = ±1 without significant contribution from other modes. Similarly, the SH vector vortex beam, as shown in Fig. 4(c), has major contribution from the OAM mode of $l$ = ± 2. The measured modal distribution of the higher order pump vector vortex, $|l_p|$ = 12, as shown in Fig. 4(d), confirm the purity of the vector vortex beams with major contribution from the OAM mode, $l$ = ± 12. However, the presence of other OAM modes in negligible weightage can be attributed to the inferior mode quality of the Ti:Sapphire oscillator with measured $M^2$ value of $M_x^2$ ~1.95 and $M_y^2$ ~2.65. Due to low SH conversion efficiency of the vector vortex beam at higher orders, we could not measure the modal distribution of the SH beams of higher orders.

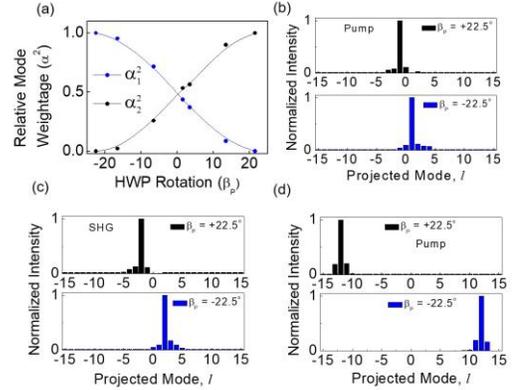

Fig. 4. (a) Experimental (dots) and theoretical (solid lines) variation of relative weightage of the orthogonally polarized modes of the SH vortex beam of order, $|l_{sh}|$= 2 as a function of the half rotation of the HWP2. Modal distribution of vector vortex beams of (b) pump of order, $|l_p|$=1, corresponding (c) SH beam of order, $|l_{sh}|$=2, and (d) the higher order pump, $|l_p|$=12.

We have studied the single-pass frequency-doubling characteristics of the vector vortex beams. Keeping pump power constant at 860 mW, we have measured the SH power of the pump vector vortex beam of different orders, $|l_p|$= 1, 2 and 3, focused using lenses of different focal lengths. For comparison, we have also studied the focusing dependent SH power of the Gaussian pump beam, $l_p$ = 0. As evident from Fig. 5(a), like Gaussian pump beam, the SH power of vector vortex beams of orders, $l_p$ = 1, 2, and 3 increases from 6.8, 2.9, 0.9 mW to 20.5, 11.6, 6.82 mW, respectively, with tight focusing arising from the decrease of focal length of the lens (L1) from $f$ = 100 to 25 mm. Although further tighter focusing can increase the vector vortex SH power, the spatial walk-off effects of the crystals split the higher-order vortex into single charged vortex beams at very tight focusing [25]. Using the lens of focal length, $f$ = 25 mm, we have measured the single-pass SHG efficiency of the vector vortex beams of different orders at constant pump power of 860 mW. The single-pass SHG efficiency of the vector vortex beam, as evident from Fig. 5(b), decreases from 6.3% for $l_p$ =0 (Gaussian) to 0.8 % for $|l_p|$=3. Such decrease in SHG power with order can be attributed to the increase of dark core size (area) of the vector vortex beam with the order as shown in the inset of Fig. 5(b).

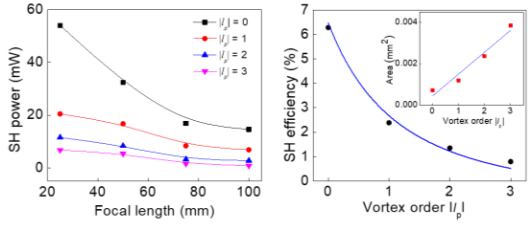

Fig. 5. (a) Dependent SH efficiency of vector-vortex beams of orders, $|l_p|$ = 0 (Gaussian), 1, 2, 3 on the focusing lens. Lines are guides to eyes. (b)Variation of SH efficiency of the vector vortex beam with its order. (Inset) Dependence of vortex area on its order. Solid lines are fit to the experimental results.

We pumped the dual-crystal with vector vortex beam of order, $|l_p|$ = 1, using the lens of focal length, $f$ = 25 mm and studied the SH power and the conversion efficiency as a function of input power. As evident from Fig. 6, the SH power increases quadratic to the pump power producing vector vortex beam with maximum output power of 20.5 mW at a pump power of 860 mW. Similarly, the single-pass SHG efficiency increases linearly with the input pump power resulting a maximum single-pass efficiency of 2.4% without any sign of saturation effect. Therefore, one can expect increase in the frequency-doubled vector vortex beam power with further increase in the pump power. The absence of saturation effect in single-pass SH power of the vector vortex beam is also evident from the linear variation of SH power with the square of the pump power as shown by the inset of Fig. 6. We have also measured the peak-to-peak power fluctuation of the SH vector vortex beam to be 10 % over 1 hour similar to the power fluctuation of pump laser.

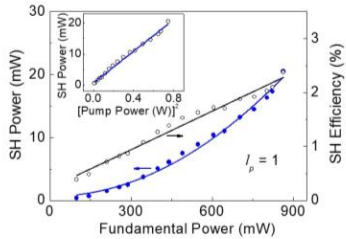

Fig. 6. Variation of SH power and efficiency of pump vector vortex order, $|l_p|$ = 1 as function of pump power. (Inset): Dependence of vector vortex SH power with the square of pump power. Solid lines are fit to the experimental results.

Using an intensity autocorrelator (pulsecheck, APE) and a spectrometer (Ocean Optics) we have measured the temporal and spectral bandwidth of the pump vector vortex beam of all orders to be 100 *fs* (after the chirped mirrors) and 55.4 nm centred at 810 nm, respectively. The broadening in the pump pulse width can be attributed to the dispersion of the optical components used in the experiment. Similarly, the spectral width of the SH vector vortex beam was measured to be 2.7 nm closely matching with the calculated spectral acceptance bandwidth of 0.6 mm long BIBO crystal [26]. The lower acceptance bandwidth of birefringent crystals can be the major constrain for lower single-pass SHG efficiency of the vector vortex beams. Due to the unavailability of suitable crystals and optics of the auto-correlator, we could not measure the temporal width of the frequency-doubled vector vortex beams.

Future projects in the fields of classical and quantum optics demand high power vector vortex beams with large OAM across the electromagnetic spectrum at ultrafast time scales. Conventional techniques of using SLMs to generate vector vortex beams have limitations due to low damage threshold and restricted spectral coverage. Here, we have demonstrated the first nonlinear generation of ultrafast higher order vector vortex beams through single-pass SHG process. Using type-I phase matched, dual crystal scheme, we have generated higher-order Poincaré beams with vortex orders up to $|l_{sh}|$ = 24 with output power maximum of 20.5 mW at a single-pass conversion efficiency as high as 2.4%. We have observed the focusing dependence of the vector vortex beam for optimum single-pass conversion efficiency. We have also studied the control in the weightage of the modes of the vector vortex beam for possible transformation of vector vortex beam into scalar vortex beam by changing the polarization state of the pump beam. The measurement on modal distribution confirms the generation of high purity vector vortex modes in both pump and SH wavelengths.